\documentclass[prl,a4]{revtex4}
\usepackage{graphicx}
\begin{document}
\title{Standard Model Masses and Models of Nuclei}
\author{Alejandro Rivero}
\affiliation{EUPT (Universidad de Zaragoza), 44003 Teruel, Spain}
 
\pacs{21.30.-x, 14.80.Bn}
\keywords{nuclear mass, Higgs mass}



\begin{abstract}
In nuclear levels, the subshells responsible for doubly magic numbers happen
to start their filling with nuclei having the same mass that relevant
Standard Model bosons.
Thus, via an undetermined many body effect, these bosons could actually
contribute to the nuclear force.
\end{abstract}

\maketitle

\section{Introduction}

\begin{figure}
    \centering
    \includegraphics[width=9cm]{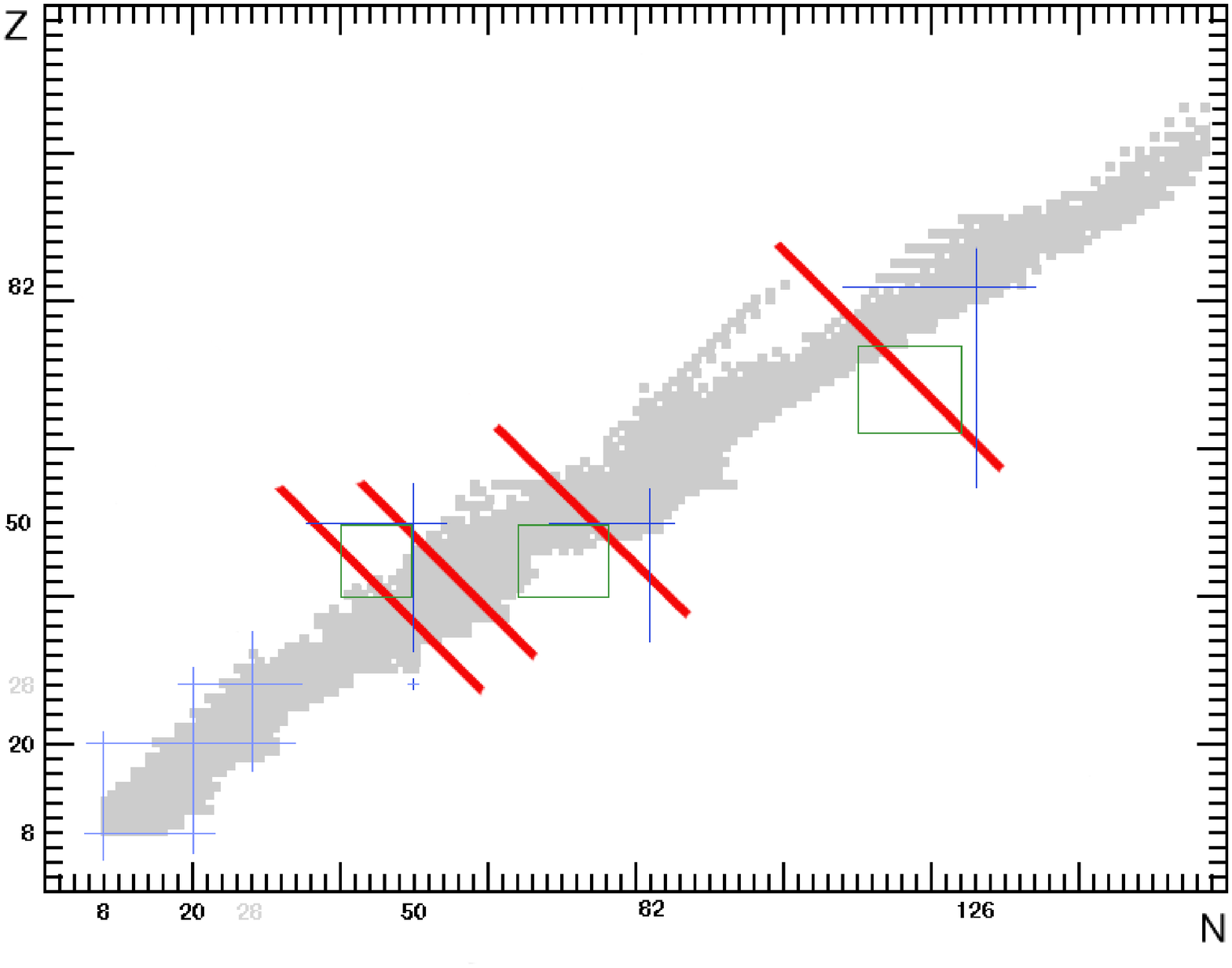}
    \caption{Crosses mark doubly magic numbers. Diagonal isobars correspond
    to the masses of W,Z, H \cite{ellis} and Top. Rectangles mark
    zones of orbital filling, with filling order from \cite{k},
    for the spin-orbit depressed subshells. }
    \label{PN}
\end{figure}

The measurement of the highest masses in the standard model, namely those
of the top quark and the particles W and Z, shows an intriguing coincidence:
the mass of the top occurs slightly before than the mass of the doubly
magic nucleus of 208 Pb, while the mass of Z and W occur slightly before
of the mass of 100 Sn.

These two magic numbers need of an abnormally high spin-orbit splitting, and there
is only a third doubly magic number sharing this need: the one of 132 Sn. Then, the
ALEPH preliminary measurement \cite{ellis} of a mass of 115 GeV for the Higgs boson,
just slightly before the mass of this Tin isotope, makes imperative to report this triple coincidence
 and to state the problem, which I do in this letter. Due to the
continuous action of strong nuclear forces, the weak bonding approximation is
not valid in this context, so it is plausible that we are seeing some kind of
general mesoscopic effect coming from many-body quantum mechanics, and not just a
specific effect of nuclear physics.

It is known that magic numbers in nuclei are generated by
separating levels via spin-orbit coupling.  Low numbers, as the 28 or 20
in doubly magic Ca48 or Ni56, can be fit using
traditional potentials with a
relatively weak spin-orbit splitting, if any. Numbers
50, 82 and 126 are got because the respective
subshells 5g9/2, 6h11/2 and 7i13/2 get large gaps and fall
into the lower shell.
Traditionally this was obtained via a purely phenomenological
Hamiltonian, consisting of an empirical spin-orbit coupling
and a Nilsson term $+k_n j^2$ depending of the shell. Modernly
a relativistic approach based in meson exchange lets one
to derive the spin orbit term from first principles. Still,
the models are restricted to low nucleon numbers, because of
computational effort. Some high (N,Z) models have been essayed,
but they fail to reproduce the phenomena of double magicity and
its associated quenching.

Here we see that some empirical coincidences are related
to the availability of virtual J=0 and J=1 particles from the Standard Model.
They seem to imply a need to correct meson exchange by adding the
whole set of standard model interacting particles. Such particles should
be noticeable during the process of filling the subshells responsable
of the magicity, thus they occur slightly before of the double
magic numbers.

Let me to organise the letter in small sections, so I will first
comment on global issues, then to look into each level in
detail, and finally to do some remarks about their possible
significance.

\section{mass landscape}

There are four highly massive particles in the standard model: the
vector bosons W,Z, the scalar Higgs boson, and the Top quark. This
one can not appear free but in composites, the simplest being
the bosons $(t \bar u), (t \bar d)$, etc, all of them having
a mass near to the Top mass. Thus we will refer to all these
particles as "the standard model bosons".


If we draw the traditional plot
of energy per nucleon, for instance for the Audi-Wapstra
experimental tables, we notice three peaks\footnote{about or less
than 1 percent in magnitude.} in the
descending part. Converting from GeV to atomic mass units, we
can draw over the plot four lines indicating the respective
masses of W, Z,Higgs (conjectured by LEP-2), and Top.
Then we note that boson masses happen slightly before the
peaks.

 Now, the peaks in such plot are mostly due to the neutron magic
numbers, but if one remembers that inside each neutron line
50, 82, 126 there is a doubly magic number, then it is more
sensible to examine the proximity of the SM masses to the
doubly magic numbers. This is done in figure \ref{PN}, by
using the traditional N-P plot of nuclides. Here we plot
diagonal isobaric lines for each massive particle, and
we see that effectively the lines happen near the
crosses corresponding to doubly magic numbers.

In order to determine how near is "near", we need to
add some information to the plot. Namely, we will
consider the subshell structure before each double
magic number.

Also, as our masses are incorporated directly from the standard
model, without direct relation to a nuclear model, we could
expect to find some signal of them in mass systematics.
Indeed the droplet model FRDM 1992 \cite{droplet},
which is the state-of-the-art
mic-mac model, shows strongly this signal for the W,Z, and a
weaker error signal for higgs and top, where a quadrupole
correction dominates the adjust.



a)P=82, N=126

The mass of the top quark was measured by the Tevatron in
1994, and it amounts, by direct observation, to 174.3 GeV,
about 187.1 amu.

\begin{figure}
    \centering
    \includegraphics[width=7cm]{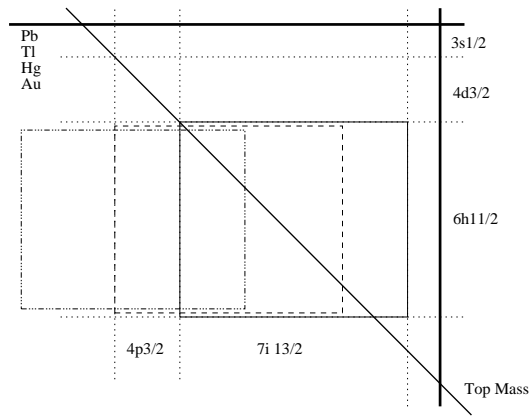}
    \caption{Diagonal isobar signals the mass of the top quark. Levels and
    solid rectangle according scheme from \cite{k}. Other possibilities are drawn
    as dashed rectangles}
    \label{top}
\end{figure}

The proton magic number 82 is due to the 6h11/2 subshell, which
actually is under two other levels, the 3s1/2 and the 4d3/2.
The neutron magic number 126 is due to subshell 7i13/2. If
we use Klinkenberg 1952 filling scheme \cite{k} to draw the rectangle
 of nuclei with partial fillings of
(6h11/2,7i13/2), the diagonal approaches closely to the
$t$ quark isobaric line. The situation
is shown in figure \ref{top}.

We should alert the reader that the filling scheme is not well
defined because the energy difference between odd and even
number of nucleons is enough to alter the energy levels. Thus
if we get the
shell scheme for 208Pb from Bohr-Mottelson 1969\footnote{indirectly quoted
in \cite{goriely} this year 2003, so it should be still considered
state-of-the-art!}, then the levels 5f5/2 and
4p3/2 seem to raise above the 7i13/2, and then the
isobar line just cuts a corner of the rectangle.
This kind of complications are usual
and we will find the same issue in the N=82 range.

One should consider also that a 5 GeV range is available starting
from the top mass, if we want to consider here all the
"mesons" from $t \bar d$ to $t \bar b$. Of course, the highly
unstable $t \bar t$ "meson" has a mass far away from the nuclear
data.

b)P=50, N=50

Here we still have the same subshell for protons and neutrons,
namely the 5g9/2. Thus the relevant nuclei are in a square.

The masses of W and Z were measured by CERN collaborations
in 1982-3. They amount to 80.423 and 91.1876 GeV,
resp. 86.3 nd 97.9 amu.

While both masses are inside the square bracketed by the subshells,
some models could also be interested in its average, 92.1 amu, which roughly
closes the diagonal of the square.

\begin{figure}
    \centering
    \includegraphics[width=4.5cm]{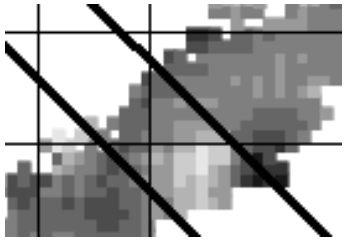}
    \caption{Horizonal (vertical) lines mark start and end of the 5g9/2 subshell
    for protons (neutrons). Thus upper-right corner is doubly magic 50,50.
    Diagonal isobars are drawn at masses of W and Z. The
    background shows FRDM-1992 mass error \cite{droplet} in this area of the
    nuclide table being black +1.5 MeV, white -2.0 MeV }
    \label{WZ}
\end{figure}

If one draws the mass lines over the plot of errors in the
microscopic-macroscopic mass formula FRDM-1992\cite{droplet}, it can be noticed
that outside from the square the lines coincide with an huge error area
in the model. In figure \ref{WZ} we have included this data.

c)P=50, N=82

Even without the higgs, the two previous coincidences should be enough
to consider an extension of nuclear models. Now,
a Higgs-like event was reported by LEP-2 collaborations in their final
year 2000 run, with a mass of 115 GeV; roughly 123.46 amu. We can
plot it and see how near it is of the extant doubly magic number.


The magic number P=50 is due to the subshell 5g9/2 as in the previous
case. The magic number N=82 is due to subshell 6h11/2. This should
do a $10\times12$ rectangle.

An unexpected complication is the overlapping of neutron subshells 6h11/2,
3s1/2 and 4d3/2. Single nucleons prefer one subshell, while paired
nucleons prefer another. It is difficult to establish the right order,
if it exists. Even when the 6h11/2 subshell is supposed to be under the
other two (or three!), the first excited level is always provided by it.

\begin{figure}
    \centering
    \includegraphics[width=5cm]{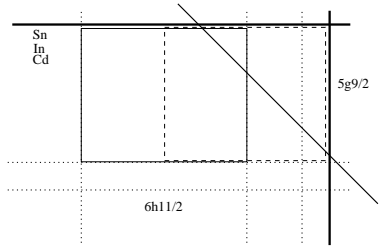}
    \caption{Diagonal isobar marks the conjectured mass of the SM Higgs scalar.
    solid rectangle takes subshell filling order from \cite{k}, dashed rectangle
    takes subshell ordering from HFB models.}
    \label{H}
\end{figure}

If we ignore the overlap and we
we accept the old criteria from Klinkenberg, putting the 6h subshell as
the lower one, then the LEP-2 event simply cuts a corner of the rectangle.
The same discrepancy, at a smaller scale, happens if we take Bohr-Mattelson.
And if we draw a whole elongated rectangle accounting for all the
possible occupations, then it is not so clear where should we draw the diagonal.
In any case, we have reflected all these possibilities in figure \ref{H}.

On other
hand, we should point out that all the modern models based on HFB mass
formulae fail to reproduce the
narrow overlap and the priority of 4d3/2, and they assign the highest
energy level to the 6h11/2 subshell,
which then closes the whole shell (check the plots and data in
\cite{goriely}). This should be the best approach, as then the
$10\times 12$ rectangle is directly attached to the doubly magic corner.

The area near this mass isobar has been studied also beyond Z=50, in the
context of Chiral Doublet Structures \cite{ch}

\section{Analysis and Remarks}

Our empirical approach is partly muddled because of the proximity between
different subshell splittings. We refer the reader again to
\cite{k} to get an idea of the complexity of these determinations.
On the other hand, it could be that all the levels are actually
trying to compete to fit in the isobaric lines.

While it is possible to believe that this effect is just a random
numerological coincidence, it covers all the cases where the naive
shells plus spin-orbit do not suffice to justify the observed magicity,
and specially it suggests that a motivation can exist for the effect
of double magicity. There are not more highly massive Standard Model particles,
and there are no more double magic nuclei with high spin-orbit splitting.
It could be argued that the 28 shell is not
included in the list, but this shell needs only a mild spin-orbit
split, at the reach of classical models. It is very unusual to find
such level of agreement in numerology with congruent units: most
numeric coincidences appear in adimensional considerations, or when
comparing numbers coming from different systems of units.

The enhancement happens in the depressed subshells, which agrees with
the phenomenological approach that enhances the spin-orbit separation
by incorporating a dependence on angular momenta.
Still, We can not, at this moment, provide an explanation of the effect. The
simultaneous fit of both proton and neutron subshells suggest that
the effect includes an isospin symmetry. 

For the top quark mesons, both J=1 and J=0 particles are available. More
specifically\footnote{thanks to P. Walters for this remark} the pseudoscalar
0-+ should be expected with more probability that the scalar 0++, while the
vector 1-{}- is always "the backbone of any meson spectroscopy". For the
W and Z, we should look for axial effects influencing angular momenta, and also
for other combinations explaining the role of W and Z in the error data of
macroscopic models.
If the LEP-2 event is confirmed to be the Higgs scalar particle, then
we should look for couplings to J=0 scalar.
It should be remembered that $J=0$ coupling is the trademark of the
scalar $\sigma$-meson in relativistic models of the nucleus.
The status of this $\sigma$ meson is experimentally troubled. It is
usually assimilated to a $f_0(600)$ boson (see \cite[pg 450]{pdg}) by the
HEP community, but nuclear authors prefer to remark that "there is
no experimental evidence for a free $\sigma$ meson, although the $\sigma$
field is a crucial ingredient of relativistic mean-field models" \cite{scmf}.
Usually the $\sigma$ scalar is seen as a two-pion particle, ie its scalar
nature comes from two identical pseudoscalar couplings.

Our particles directly from the Standard Model could interfere with the effect of
the $\sigma$. It could happen either as an small modification of the
$1/m_\sigma$ mass term, or as a direct interaction with the nucleon
density. It seems from our data that the meson exchange interaction is enhanced
when the mass of the whole ensemble $N+Z$ of nucleons equals the mass of some
available SM boson. This could be more explicit in the second case.

Also, it could
be possible to develop some influence in the effective nucleon mass, which,
at the end, is the main responsible for the spin-orbit coupling.

On other hand we could be seeing a quantum many-body effect. The
interaction propagated via a massive particle will depend both on the mass of
the nucleon, $M$, and the mass $m$ of the boson. But there is also a multiplicative
factor depending of the number of nucleons, i.e. of $M_n/M$, being $M_n$ the mass
of the nucleus. So the total interaction can be written in terms of $(M_n,M,m)$ and
we could expect some surprise\footnote{An intermediate suggestion -to be tested- is
to calculate the interaction between the whole neutron and proton balls, with
respective masses $M_N,M_P$, when the interacting particle has $m\approx M_N+M_P$.
This is suggested because observation of quenching hints that both subshells
must be present for the double magicity phenomena to appear.}
when $M_n\approx m$. Besides, $M_n$ could be directly the relevant quantity
for kinematics.

The strong nuclear force, as driven by the usual system of meson
exchanges, provides a kinematical glue because our forces are small
compared to the coupling constant from pions. We are working with
bound states; any momenta exchanged between a bound particle and the
nucleus is shared between all the particles of the nucleus before the 
"orbiting" particle gets able to turn around and interact from the opposite 
direction. So the acceleration from this momentum exchange depends on 
the mass of the whole nucleus, and this whole mass is more relevant
that the mass of the nucleon -or quark- actually responsible of the exchange. It
happens also,  in some classical inelastic collisions, that the conversion of 
energies $\mbox{\it collision}\rightarrow
\mbox{\it internal} + \mbox{\it mass center}$ depends on the coupling constant
of the internal force. The perturbation is to be measurable because in a yukawian
potential it depends on the product of mass (inverse range) times coupling.

In any case, if the recoil depends of the whole mass, it is difficult to justify why
should one to cut-off the calculation at the scale of the single nucleon.
Regretly, most meson theories  are effective QFT with a cutoff $\Lambda <$ 1 GeV,
so they are not directly suitable to incorporate the highest particles.

 With a clear
background, and in the absence of serious Effective Field Theory
motivations, the ratio $\Gamma/m$ between pole width and
pole mass of a given meson should be a consideration more fundamental
than simply to look if such mass is above or below the one of the
proton. It should be of some help if similar effects were searched in
few nucleon systems. The scale of J/psi, B and Upsilon could
be an extra contribution respectively related to the variations of mass
at 4He, 8B, 12C, or to contribute to the valleys between, or simply to be
hidden under other effects. Furthermore, perhaps
here the nucleon number is too small or the background of other particles
too big. So the most we can say is that the low mass range does not contradict
our observation, but that it is not a definitive support. Ab-initio models could
be able to fit the data, in the future, without requiring the mildly massive
mesons.

Beyond theoretical frameworks, another clue to this effect can arrive from an
analysis of the errors in current models of masses. Anomalies are to be
expected near mass numbers 86, 98, 123, 187 (W, Z, Higgs, tops). We have remarked
this point in figure \ref{WZ}. Indeed the error analysis of \cite{mexico} for the same 
model hints the same results, and also the variations between theoretical and
real error in fig 7 of  \cite{droplet}. Besides, the recent analysis of Huertas for FST model
shows slope changes in the error at this mass numbers, see figure 4 of
\cite{huertas1} and fig. 13 of \cite{huertas2}. 

Of course it shall not escape to any reader that a Mic-Mac model including high
SM masses as free parameters could be used to predict these SM values, via
least squares or any other variational fit.  For instance one can calculate the 
model parameters dependence on the masses of the nuclei used in the fit; 
say $g_>(m)$ fitting to all nuclei greater than $m$, or $g_<(m)$ fitting for
all the nuclei with mass smaller than $m$, or even some intermediate 
fit $g_{+10}(m)$. Then any sudden jump in $g(m)$ could be used to
signal a unexpected effect around mass $m$.

Our work is in some sense falsifiable at mid term: If either no particle appears near 115 GeV or QHD proves able to reproduce the spin couplings for double magicity, the coincidence will be invalid in the first case, accidental in the later. A ten-years
lapse will do the final judgement, as computing power increases and the LHC does
its hunt. 



The author wants to acknowledge L.J.Boya and other members of the Theoretical Physics Department at Zaragoza for discussions, as well to P. Walters at the internet
physicsforums \cite{pforums}.


\section{appendix}

As of today, the previous comment has been rejected from practically
all the letter journals of its thematic\footnote{see my webpage for
details on this mundane thing}. The rejection has been mostly at
editorial level, so I have not got any useful input from referees
nor pairs. 

During this time I have mostly tried some theoretical preparation, but I can
not report of serious advance in this side. On the other side, the
empirical, we are luckier, perhaps. I have mentioned in the paper that
we should expect huge error peaks in mass models that are unable to make
place for the particles we are looking for. Indeed we have shown this
effect for W and Z in an area of the FRDM model. But this model does not
show huge discrepancies for 115 and top. Do our conjecture fails or,
more probably, the model is able to adjust itself to nullify the
discrepancy? Last week I did some plottings that hint towards the
second possibility.

\begin{figure}
    \centering
    \includegraphics{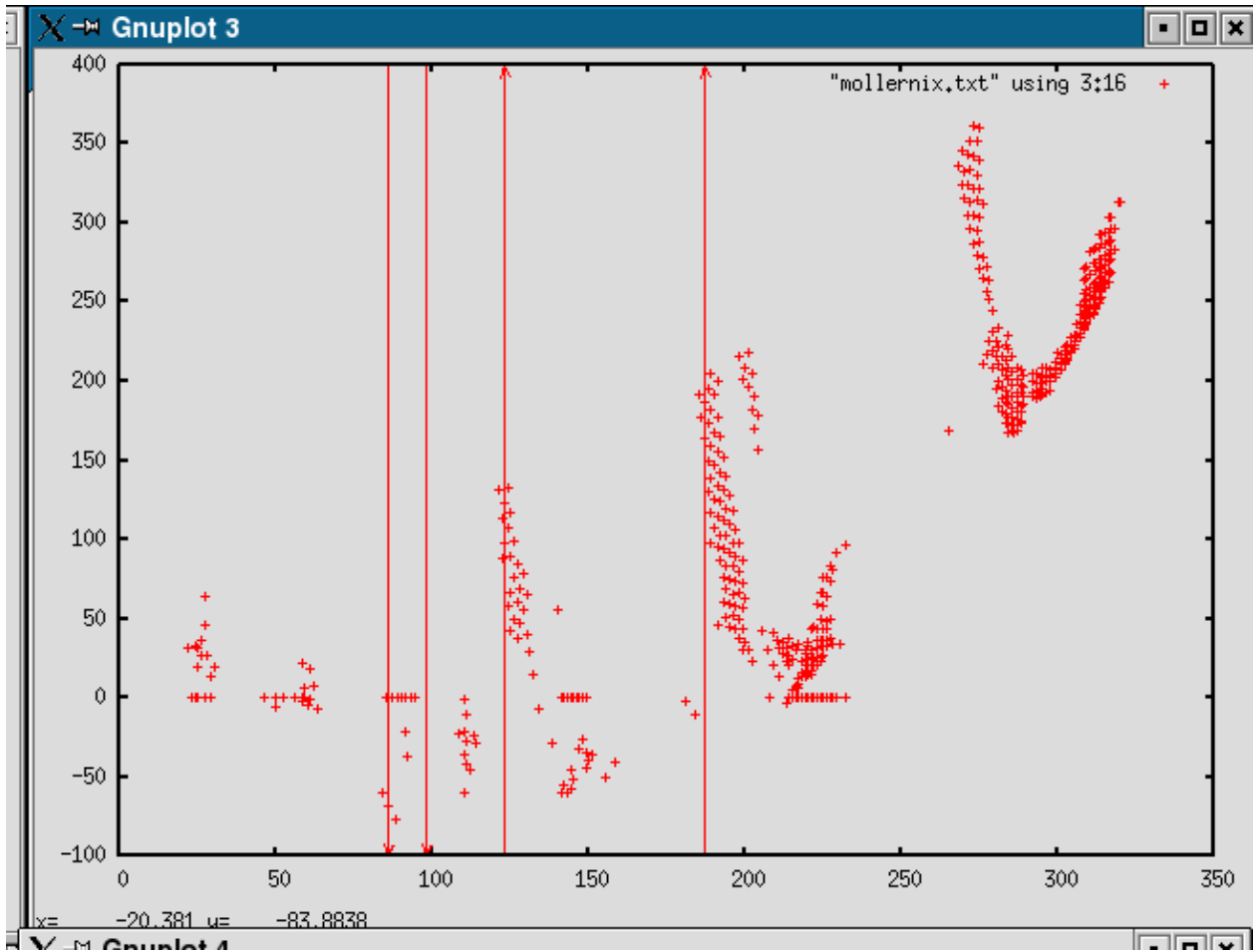}
    \caption{By mistake, I plotted a mixed graph of FRDM data... }
    \label{mistake}
\end{figure}

I took upon myself the task of plotting error graphs for some tables
available in the net. I started with the FRDM table, and a mistake
when manipulating the data in gnuplot produced the result you can see
in figure \ref{mistake}. 

\begin{figure}
    \centering
    \includegraphics[width=9cm]{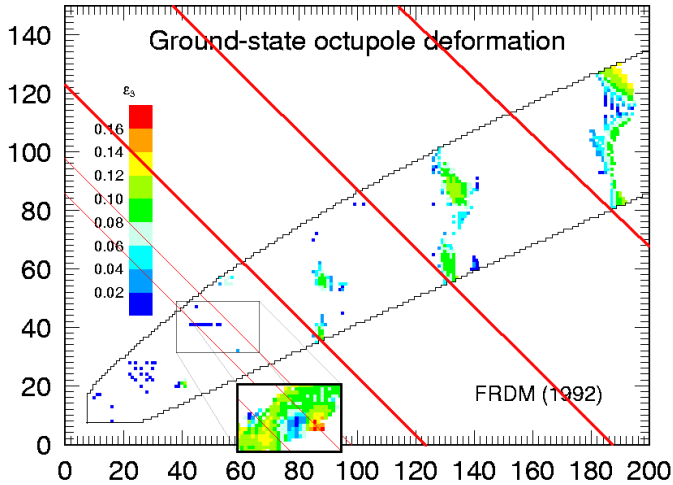}
    \caption{... and when rightly plotted, a hint to the neutron dripline appears. This
    is figure 10 of \cite{droplet}, plus an inset of figure 20.  The only addition to the 
    original plot are the diagonal isobars, at $M_W$, $M_Z$, 115 Gev, 
    $M_t$ and 246 Gev.}
    \label{e3}
\end{figure}

I was mixing different column data in the same plot, but still getting
an intriguing result. By examining the data, it happened that I was
plotting a conditional: for some nuclei the model uses a kind of
shape correction, for some others it chooses another. I went to the
original paper \cite{droplet} to check the selection of nuclei, and I found the 
coincidences shown in figure \ref{e3}

\begin{figure}
    \centering
   \includegraphics[width=10cm]{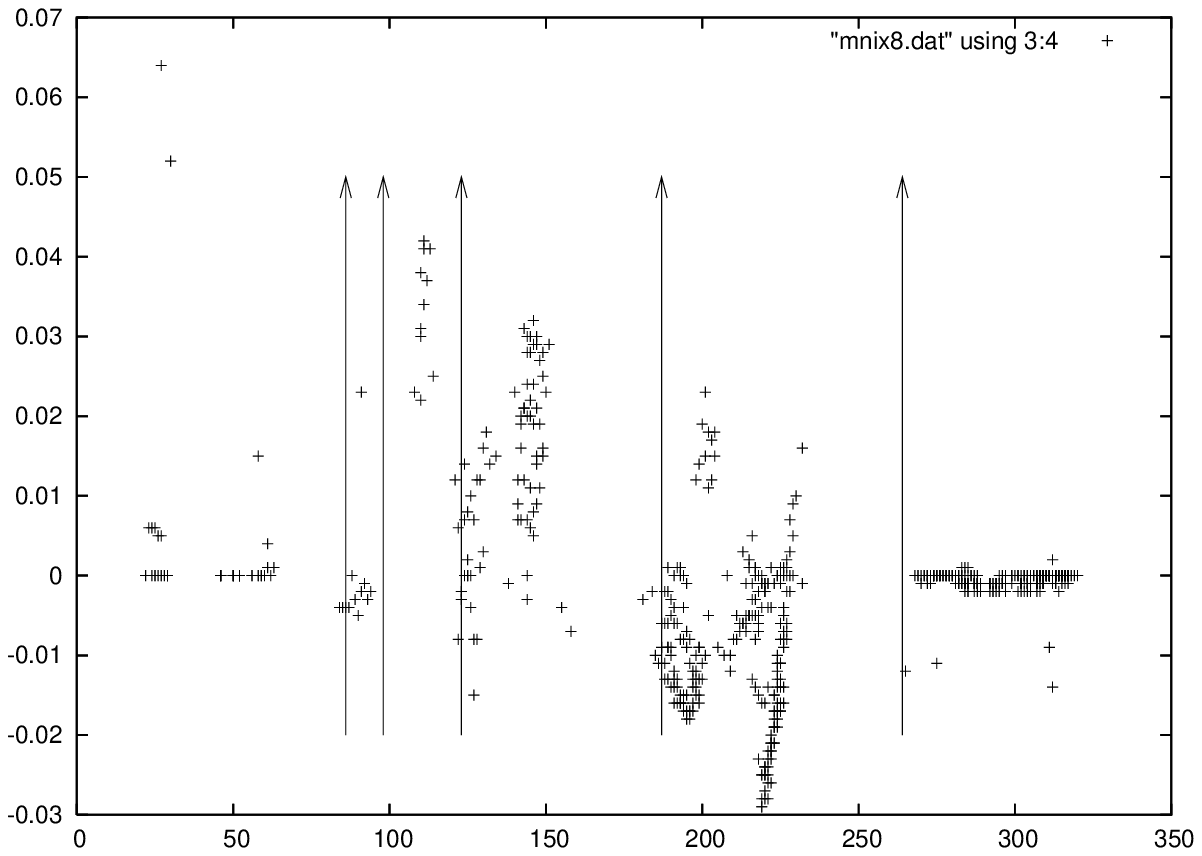}
    \caption{values of $\epsilon_6$ discarded in the FRDM when using
    instead the deformation parameter $\epsilon_3$. Note the qualitative
    difference in the signal of electroweak vacuum.}
    \label{e6sym}
\end{figure}

\begin{figure}
    \centering
   \includegraphics[width=9cm]{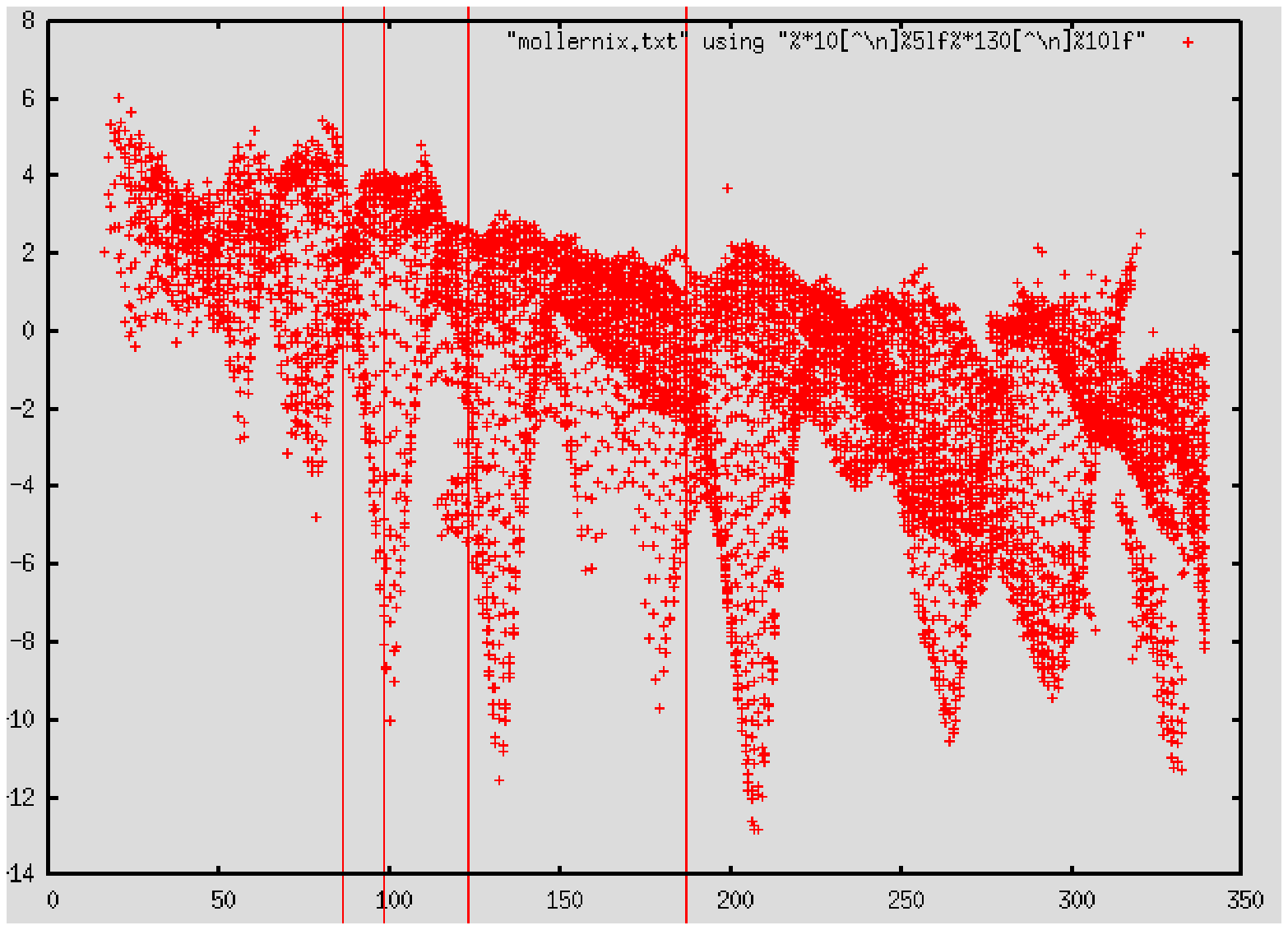}
    \caption{plot of the calculated microscopic correction of FRLDM \cite{droplet}, from
    table available at \cite{tablas}.}
    \label{micr}
\end{figure}

The apparition of the electroweak vacuum (the expected value of the
higgs field) is surprising because it does not imply a standard
model particle. The discarded $\epsilon_6$ in this area is qualitatively
different from the others (fig. \ref{e6sym}). Besides, the 246 Gev range is
also noticeable in the -very noisy- plot of calculated microscopic energies for
the twin model FRLDM. 
(figure \ref{micr}). 

The emerging role of the neutron drip line is consistent with out
thinking of interpreting the rest of the nucleus as a single 
particle: at the dripline, the neutron skin orbits the
rest of the nucleus in a dense version of the halo of light drip line
nuclei.

We see that this concrete model uses shape corrections to fill the
expected discrepancies. But what about other models?  Results vary. 
Generically, when the error plot has noticeable peaks, the pattern
of electroweak scales appears. By looking at \cite{otros}, about two thirds
of the models present very noisy error patterns, but still we have another
third showing curious peaks at the appropriate ranges. 

\begin{figure}
    \centering
   \includegraphics[width=18cm]{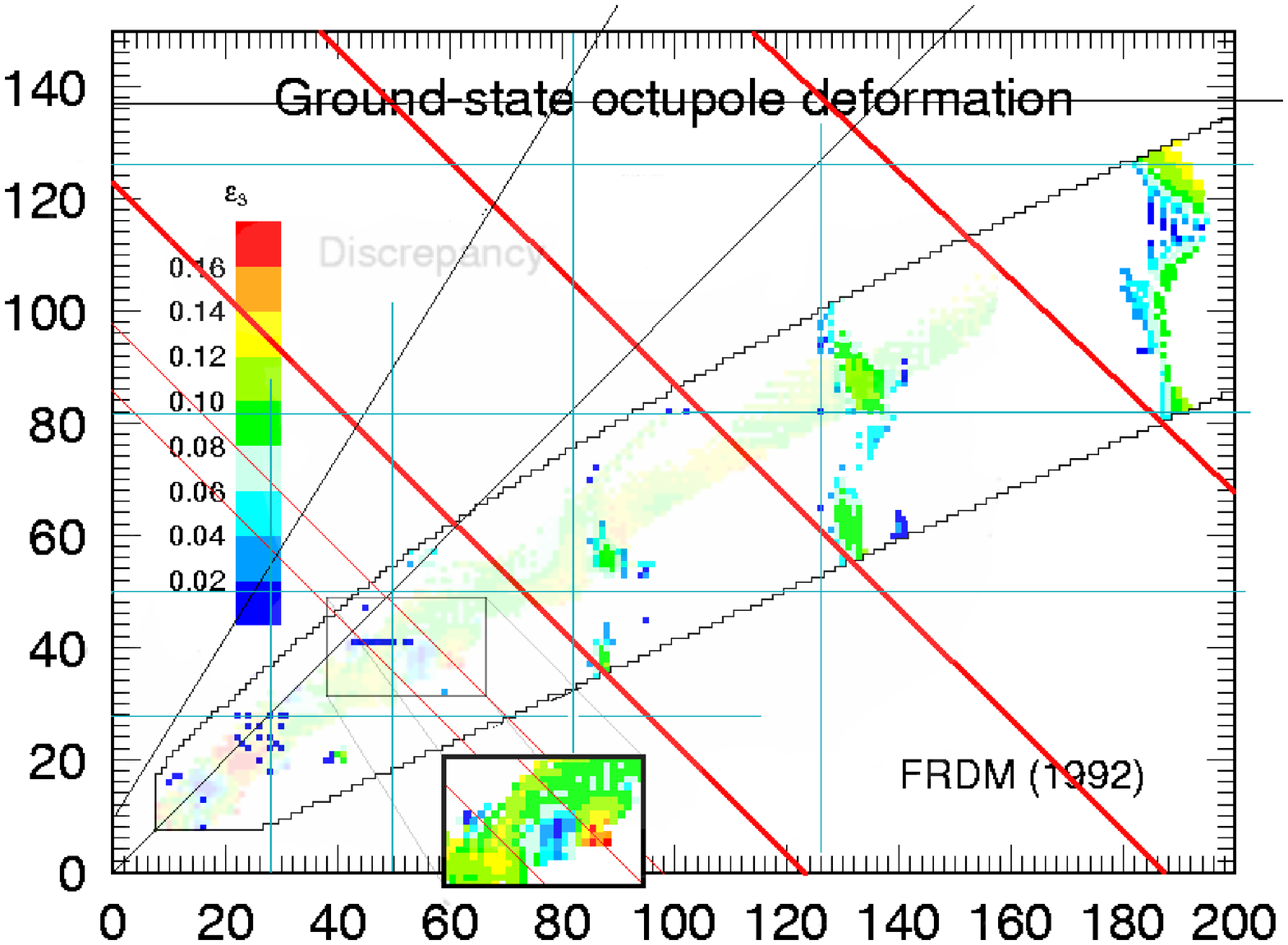}
    \caption{Same as \ref{e3} but with some additional support  lines and data}
\end{figure}


\begin{thebibliography}{30}

\bibitem{scmf}
M. Bender, P-H. Heenen, P-G. Reinhard
\textit{Self-consistent mean-field models for nuclear structure}
Rev. Mod. Phys. 75, 121 (2003)
http://link.aps.org/abstract/RMP/v75/p121

\bibitem {cohen} B. L. Cohen
{\it Observations on the Energies of Single-Particle Neutron States}
Phys. Rev. 130, 227-233 (1963)
 http://link.aps.org/abstract/PR/v130/p227

\bibitem {ellis}
J. R. Ellis
{\it The 115 GeV Higgs Odyssey}
http://arxiv.org/abs/hep-ex/0011086

\bibitem{tablas}
R.B. Firestone, 
{\it LBNL Isotopes Project Nuclear Data Dissemination Home Page}. 
Retrieved May 2003, from http://ie.lbl.gov/toi.html. 
and http://ie.lbl.gov/toimass.html

\bibitem{goriely} S. Goriely et al., {\it Further explorations...}, Phys.
 Rev. C 68, 054325 (2003).

\bibitem{mexico}
Jorge G. Hirsch, Alejandro Frank, Victor Velazquez
 {\it Residual correlations in liquid drop mass calculations}
 Phys.Rev. C69 (2004) 037304
 http://arxiv.org/abs/nucl-th/0306049
 
\bibitem{huertas1}   
Marco A. Huertas 
 {\it Effective lagrangian approach to structure of selected nuclei far from stability}
 Phys.Rev. C66 (2002) 024318; C67 (2003) 019901
 http://arxiv.org/abs/nucl-th/0203019

 \bibitem{huertas2}     
Marco A. Huertas  
 {\it Extension of effective lagrangian approach to structure of selected nuclei far from stability} 
 Acta Phys. Pol. B 35, 837 (2004)
 http://arxiv.org/abs/nucl-th/0304016

\bibitem{k} P. F. A. Klinkenberg
 {\it Tables of Nuclear Shell Structure}
 Rev. Mod. Phys. 24, 63-73 (1952)
 http://link.aps.org/abstract/RMP/v24/p63

\bibitem{trends}
D. Lunney, J. M. Pearson,  C. Thibault
{\it Recent trends in the determination of nuclear masses}
Rev. Mod. Phys. 75, 1021 (2003)
http://link.aps.org/abstract/RMP/v75/p1021


\bibitem{droplet}
P. M\"oller, J. R. Nix, W. D. Myers, W. J. Swiatecki
 {\it Nuclear Ground-State Masses and Deformations}
 Atom.Data Nucl.Data Tabl. 59, 185-381 (1995)
 http://arxiv.org/abs/nucl-th/9308022

\bibitem{pdg} Particle Data Group, {\it Review of Particle Physics},
Phys. Rev. D 66 (1-I), p. 1 (2002)  http://pdg.lbl.gov/

\bibitem{pforums}
Vv.aa., {\it Physics/ Nuclei\&Particles/ living without CERN},
at {\tt www.physicsforums.com}, threadid = 9824

\bibitem{ch}
 K. Starosta {et al.},
{\it Chiral Doublet Structures in Odd-Odd N   =   75 Isotones: Chiral Vibrations}
Phys. Rev. Lett. 86 (6) 971-974 (2001)
http://link.aps.org/abstract/PRL/v86/p971

\bibitem{otros}
VV. AA., 
Atom.Data Nucl.Data Tabl. 39, 185 (1988)
see also \cite{tablas}










\end{thebibliography}
\end{document}